\documentclass{article}

  \usepackage{longtable}
\usepackage{amsmath}
\usepackage{array}
  \newcolumntype{L}[1]{>{\raggedright\let\newline\\\arraybackslash\hspace{0pt}}m{#1}}
  \newcolumntype{C}[1]{>{\centering\let\newline\\\arraybackslash\hspace{0pt}}m{#1}}
  \newcolumntype{R}[1]{>{\raggedleft\let\newline\\\arraybackslash\hspace{0pt}}m{#1}}
\usepackage{arydshln}
\usepackage{colortbl}
\usepackage[inline]{enumitem}
  \setenumerate{itemsep=-1pt,topsep=2pt}
\usepackage{float}
\usepackage{framed}
\usepackage[margin=1in]{geometry}
\usepackage{graphicx}
  \graphicspath{{figures/}}
  \DeclareGraphicsExtensions{.pdf,.eps,.png}
\usepackage[utf8]{inputenc}
\usepackage{multirow}
\usepackage{url}
  \makeatletter
  \g@addto@macro{\UrlBreaks}{\UrlOrds}
  \makeatother
\usepackage{xcolor}

\newcommand{\ccb}[1]{\multicolumn{1}{c}{\textbf{#1}}}	       % Cell Center Bold
\newcommand{\cdline}[1]{\cdashline{#1}[0.5pt/2pt]}		       % Columnar dashed line of 0.5pt separated by 2pt
\newcommand{\dline}{\hdashline[0.5pt/2pt]}				       % Dashed line of 0.5pt separated by 2pt
\newcommand{\etal}[1]{#1 \textit{et al.}}                      % Appends italicized et al. to the last name of an author
\newcommand{\header}[1]{
  \vspace{3pt}\noindent\textbf{#1}\vspace{3pt}
}
\newcommand{\subheader}[1]{
  \vspace{3pt}\noindent\textit{#1}\vspace{3pt}
}
\newcommand{\lland}{\ttt{AND} }                                % Logical AND to combine search strings
\newcommand{\llor}{\ttt{OR} }                                  % Logical OR to combine search strings
\newcommand{\mcb}[3]{\multicolumn{#1}{#2}{\textbf{#3}}}        % Same as \mc with bold
\newcommand{\mr}[2]{\multirow{#1}{*}{#2}}				       % Span a cell across #1 rows
\newcommand{\ttt}[1]{\texttt{#1}}						       % True Type Text
\begin{document}

% ################################################
% Top Matter

\title{\textbf{Systematization of Vulnerability Discovery Knowledge}\\{\Large Review Protocol}}
\author{
  Nuthan Munaiah and Andrew Meneely\\
  Department of Software Engineering\\
  Rochester Institute of Technology\\
  Rochester, NY 14623\\
  \ttt{\{nm6061,axmvse\}@rit.edu}
}

\maketitle

\section{Introduction}

As more aspects of our daily lives depend on technology, the software that supports this technology must be secure. We, as users, almost subconsciously assume the software we use to always be available to serve our requests while preserving the confidentiality and integrity of our information. Unfortunately, incidents involving catastrophic software vulnerabilities such as Heartbleed (in OpenSSL), Stagefright (in Android), and EternalBlue (in Windows) have made abundantly clear that software, like other engineered creations, is prone to mistakes.

Over the years, Software Engineering, as a discipline, has recognized the potential for engineers to make mistakes and has incorporated processes to prevent such mistakes from becoming exploitable vulnerabilities. Developers leverage a plethora of processes, techniques, and tools such as threat modeling, static and dynamic analyses, unit/integration/fuzz/penetration testing, and code reviews to engineer secure software. These practices, while effective at identifying vulnerabilities in software, are limited in their ability to describe the engineering failures that may have led to the introduction of vulnerabilities. Fortunately, as researchers propose empirically-validated metrics to characterize historical vulnerabilities, the factors that may have led to the introduction of vulnerabilities emerge. Developers must be made aware of these factors to help them proactively consider security implications of the code that they contribute. In other words, we want developers to think like an attacker (i.e. inculcate an \emph{attacker mindset}) to proactively discover vulnerabilities.

Over the last decade, several metrics \cite{Neuhaus2007Predicting,Neuhaus2009Beauty,Shin2008Complexity,Shin2008Empirical,Shin2011Evaluating,Shin2011Initial,Shin2011Investigating,Shin2013Traditional,Chowdhury2011Using,Meneely2009Secure,Meneely2010Strengthening,Meneely2013Patch,Zimmermann2010Searching,Scandariato2014Predicting,Walden2014Predicting,Perl2015Finding,Younis2016Fear} have been proposed to assist developers in discovering vulnerabilities. However, the adoption of these metrics in mainstream software engineering has been limited owing to concerns such as the high frequency of false positive from predictive models that use the metrics as features, the granularity at which the metrics operate, and lack of interpretable and actionable intelligence from the metrics \cite{Morrison2015Challenges}. These concerns may be mitigated by considering the empirical properties that the metrics embody beyond their predictive capabilities. We must consider the metrics not as mere features in a vulnerability prediction model but as agents of feedback on security. In other words, we must ask ourselves \emph{what is the metric telling us?} and \emph{what can we ask developers to do?} In essence, we must \emph{humanize} the metrics such that they can communicate the rationale for a source code entity to be considered vulnerable. In interpreting the feedback that a metric provides, developers gain awareness of the potential factors that lead to vulnerabilities thus aiding in inculcating an attacker mindset.

The goal of our research is \emph{to assist developers in engineering secure software by providing a technique that generates scientific, interpretable, and actionable feedback on security as the software evolves}. The first step toward the accomplishment of this research goal is the systematization of the state of the art in vulnerability discovery knowledge, specifically, in the realm of metrics-based discovery of vulnerabilities.

In this report, we describe the review protocol that will guide the systematic review of the literature in metrics-based discovery of vulnerabilities. The protocol have been developed in adherence with the guidelines for performing Systematic Literature Reviews in Software Engineering prescribed by Kitchenham and Charters \cite{Kitchenham2007Guidelines}. In addition to the guidelines prescribed by Kitchenham and Charters \cite{Kitchenham2007Guidelines}, specific aspects of the the review protocol have further been informed by other, more specific, guidelines. For instance, we used the guide published by Counsell \cite{Counsell1997Formulating} to aid formulating the research questions in our systematic review. We reference such specific guidelines in the sections that they are applicable in.

\section{Systematic Literature Review Process}

A typical systematic literature review is has three main stages, they are:

\begin{enumerate}
  \item Plan
  \item Conduct
  \item Report
\end{enumerate}

In each of these main stages, there are several subtasks that must be accomplished to ensure that the review is carried out in as rigorous, unbiased, and repeatable way as possible. In this report, we will describe the subtasks in the plan and conduct stages of the systematic review.

\subsection{Plan}

In this stage, the specifics of all aspects needed for the conduct of the systematic review being proposed are identified and documented. The planning stage is crucial in ensuring the rigor of the systematic review. 

The stage begins with determining if there is indeed a need for the review being proposed. The stage concludes with the publication of a review protocol (this report) to the research team. The research team will then use the protocol to guide the conduct of the systematic review, revisiting the planning stage only when there are issues identified with the protocol during the conduct stage.

In the subsections that follow, the various subtasks of the planning stage are described in detail.

\subsubsection{Need For the Review}

The literature on vulnerability discovery metrics is largely scattered across many primary studies \cite{Neuhaus2007Predicting,Neuhaus2009Beauty,Shin2008Complexity,Shin2008Empirical,Shin2011Evaluating,Shin2011Initial,Shin2011Investigating,Shin2013Traditional,Chowdhury2011Using,Meneely2009Secure,Meneely2010Strengthening,Meneely2013Patch,Zimmermann2010Searching,Scandariato2014Predicting,Walden2014Predicting,Perl2015Finding,Younis2016Fear}. To aid developers in engineering secure software, we must characterize the factors that may have led to the introduction of vulnerabilities in the past. The primary need for the systematic review (for which a review protocol is being described in this report) is to aggregate the empirical evidence demonstrating the utility of metrics to discover vulnerabilities in software. The secondary need for the systematic review is as a prelude to the subsequent research studies that will lead to the accomplishment the greater research goal of the Assisted Discovery of Software Vulnerabilities project.

As recommended by the guidelines \cite{Kitchenham2007Guidelines}, we searched for existing secondary studies addressing the same (or similar) research questions as we aim to address in our systematic review. In this search, we used Google Scholar\footnote{\url{https://scholar.google.com/}} as the query engine and \texttt{software + vulnerability + review} as the search string. While we did not find any secondary studies addressing the same research questions as we aim to address, we did find two studies (\etal{Liu} \cite{Liu2012Software} and Ghaffarian and Shahriari \cite{Ghaffarian2017Software}) that seemed related. However, these studies are not systematic review but rather surveys of the different approaches to discover vulnerabilities. The use of metrics to discover vulnerabilities in software was just one of many approaches described in the work by Ghaffarian and Shahriari \cite{Ghaffarian2017Software}.

\subsubsection{Research Question(s)}

The guidelines \cite{Kitchenham2007Guidelines} prescribe the use of the PICOC (Population, Intervention, Comparison, Outcomes, and Context) criteria to structure research questions in systematic reviews. The components (i.e. PICOC) of the research question are determined by the type of research question. The type of research question is in-turn determined by the type of evidence available in the primary studies (See Table 3 in the study by Fineout-Overholt and Johnston\cite{Overholt2005Teaching} for examples of different types of questions and the types of evidence needed to answer a given question).

In the context of our systematic review, these components of the research question are as follows.

\begin{description}
  \item[Population] \hfill
  
  The population in our review is \emph{software} since the goal is to understand the metrics that aid in discovery of vulnerabilities in software.
  \item[Intervention] \hfill
  
  The intervention in our review is the \emph{increasing or decreasing of the value of an empirically-validated metric} as a result of some activity in the Software Development Lifecycle.
  \item[Comparison] \hfill
  
  The comparison is \emph{not applicable} to our review because the goal is not to compare the outcomes from software with intervention and those from software without.
  \item[Outcomes] \hfill
  
  The outcome that concerns our review is \emph{discovery of security vulnerabilities}.
  \item[Context] \hfill
  
  In Software Engineering, the context is the context in which the primary study was conducted. The context could include factors such as where (academia or industry), what (open source or closed source), who (if applicable, practitioners, academics, or students), and how (quantitative or qualitative). In our review, the only restriction on context is that the primary study must be quantitative. By not restricting the where, what, and, who, we are likely to get a holistic perspective on the literature increasing the potential for our review to produce more generalizable conclusions.
\end{description}

\subsubsection{Protocol}

\header{Background} % -----------------------------

As mentioned earlier, the goal of our research is \emph{to assist developers in engineering secure software by providing a technique that generates scientific, interpretable, and actionable feedback on security as the software evolves}. The purpose of the systematic review (for which a review protocol is being described in this report) is twofold \begin{itemize*} \item aggregate the empirical evidence demonstrating the utility of metrics to discover vulnerabilities in software and \item assess the extent to which vulnerability discovery metrics have been validated across primary studies in the literature\end{itemize*}. When completed, the systematic review can be used to inform practitioners of the factors that have led to introduction of vulnerabilities in the past and inform researchers of the strengths and weaknesses of existing vulnerability discovery metrics.

\header{Research Questions} % -----------------------------

We will address the following primary research questions in our systematic review.

\begin{description}
  \item[RQ1 Enumeration] \hfill
  
  What \emph{metrics} have been proposed to \emph{discover security vulnerabilities} in \emph{software}?
  \item[RQ2 Validation] \hfill
  
  How have researchers evaluated the validity of the \emph{metrics} to \emph{discover security vulnerabilities} in \emph{software}?
  \item[RQ3 Applicability] \hfill
  
  How do researchers suggest the \emph{metrics} be used to \emph{discover security vulnerabilities} in \emph{software}?
  % TODO: Remove commented lines
  % How have researchers interpreted the \emph{metrics} to \emph{discover vulnerabilities} in \emph{software}? \rw{The research question is much more open ended than the other questions. Report suggested uses and then discussion the interpretability.}
\end{description}

In addition to systematically reviewing vulnerability discovery literature to address the aforementioned primary research questions, we will implement the algorithm to collect the metrics and apply the algorithm to collect the metrics from a set of subjects of studies. In implementing the algorithm to collect the vulnerability discovery metrics, and collecting them from a set of subjects of studies, we will address the following  secondary research questions.

\begin{description}
  \item[RQ4 Prerequisites] \hfill
  
  What are the prerequisites to collect \emph{metrics} to \emph{discover security vulnerabilities} in \emph{software}?
  % TODO: Remove commented lines
  % What are the challenges in collecting the \emph{metrics} to \emph{discover vulnerabilities} in \emph{software}? \rw{Are these challenges really prerequisites? Perhaps have a discussion about the subjectivity of the prerequisites being a challenge. What if a metric is only informally defined with neither a description of the approach to collect a metric nor a reference to a resource that describes the approach to collect a metric?}
  \item[RQ5 Consistency] \hfill
  
  Are the empirical properties of the \emph{metrics} to \emph{discover security vulnerabilities} consistent across \emph{software} within, and between, domains?
\end{description}

\header{Search Strategy} % -----------------------------

In the subsections that follow, the approach to identify the primary studies that will be systematically reviewed to address the research questions is outlined. The search for primary studies will be conducted in two phases: \begin{enumerate*}[label=\textbf{Phase \Roman* -}]\item \textbf{Initial (Automated) Search}: In which the search string will be used to search the various sources of candidate studies and \item \textbf{Secondary (Manual) Search}: In which the references in the candidate studies identified in Phase I will be manually analyzed to identify additional candidate studies\end{enumerate*}. There are two steps that must be accomplished before Phase I can begin, they are:

\begin{enumerate}[label=Step \arabic*]
  \item Identify sources of primary studies
  \item Identify search keywords
\end{enumerate}

A note on terminology before we proceed: the phrase \emph{candidate study} as used here is meant to refer to any study that is \emph{hypothesized} to be relevant to our review according to some predetermined criteria (i.e. is retrieved by the keywords used to search). The candidate studies are subject to thorough inclusion and exclusion criteria to identify those studies that are relevant to address the research questions posed in our systematic review. We will use the phrase \emph{primary study} to refer to a candidate study that pass the inclusion and exclusion criteria.

In both the aforementioned steps, a set of primary studies known, from manual search and prior experience, to be relevant to address the research questions posed in our systematic review is used to identify sources of primary studies and validate the search keywords. The set of known primary studies is known as the Quasi-Gold Standard Set as described by \etal{Zhang} \cite{Zhang2011Identifying}. The primary studies that compose the Quasi-Gold Standard Set in our review are listed in Appendix \ref{asec:qgsset}.

We used the Quasi-Gold Standard approach prescribed by \etal{Zhang} \cite{Zhang2011Identifying} to validate, and improve, the effectiveness of the final search string. More specifically, we use the \textit{Quasi-Sensitivity} metric defined in Equation \eqref{for:quasisensitivity}. The metric enables objective assessment of the effectiveness of the search strings. For instance, assume that the Quasi-Gold Standard Set contains 10 studies published by ACM. If a search using the ACM Digital Library retrieved 8 (of the 10) studies, then the Quasi-Sensitivity is 80\%.

\begin{equation}\label{for:quasisensitivity}
\text{Quasi-Sensitivity} = \frac{\text{\# Studies in Quasi-Gold Standard Set Retrieved}}{\text{Total \# Studies in Quasi-Gold Standard}} \%
\end{equation}

As \etal{Zhang} \cite{Zhang2011Identifying} have suggested, a rational threshold for Quasi-Sensitivity may be assumed to say that the search strategy is acceptable. We have chosen 80\% as the threshold in our review.

\subheader{Step 1: Identify Sources of Primary Studies}

We begin by identifying the sources from which our primary studies are likely to originate. Since we have a set of known primary studies (the Quasi-Gold Standard Set), we simply enumerate the publishers of these primary studies. The publishers become the sources of primary studies in our review.

The studies in our Quasi-Gold Standard Set were published by five distinct publishers: \begin{enumerate*}[label=(\arabic*)]\item ACM, \item IEEE, \item Elsevier, \item Springer, and \item USENIX\end{enumerate*}. Each of these publishers provide a service to search their respective publication databases. The name, and location (URL), of the search services used in our review are shown in Table \ref{tab:searchservices}.

\begin{table}[ht]
\renewcommand{\arraystretch}{1.3}
\centering
\caption{Search services provided by publishers of academic content.}
\label{tab:searchservices}
\begin{tabular}{ll}
  \hline
  \ccb{Publisher} & \ccb{Service (URL)}                                            \\
  \hline
  ACM             & ACM Digital Library (\url{https://dl.acm.org/})                \\ \dline
  IEEE            & IEEE Xplore (\url{https://ieeexplore.ieee.org/})               \\ \dline
  Elsevier        & ScienceDirect (\url{https://www.sciencedirect.com/})           \\ \dline
  Springer        & SpringerLink (\url{https://link.springer.com/})                \\ \dline
  USENIX          & USENIX (\url{https://www.usenix.org/publications/proceedings}) \\
  \hline
\end{tabular}
\end{table}

In addition to the search services shown in Table \ref{tab:searchservices}, we used the private search service provided by Rochester Institute of Technology (RIT), called Summon, that is used by RIT Library (\url{https://library.rit.edu/}). We chose to include Summon as a catch-all source of primary studies to include studies that may be missed by the other search services. In the past, systematic reviews have used Google Scholar (\url{https://scholar.google.com/}) for similar purposes.

\subheader{Step 2: Identify Search Keywords}

We followed the approach outlined below to identify the keywords that will be used to query sources of academic publications for candidate studies.

\begin{enumerate}
  \item We started with the population (software), intervention (metrics), and outcome (discover vulnerabilities) components from the research questions outlined earlier. In addition to these components, we also included the context (quantitative) in which primary studies should have been conducted to be considered relevant.
  \item We identified synonyms of the keywords pertaining to population, intervention, outcome, and context. We also considered truncated versions (stems) of the keywords to ensure variations in usage of the same keyword was accounted for. For instance, the truncated keyword ``vulner'' can capture both ``vulnerability'' and ``vulnerabilities''. When applicable, we also considered alternative spellings\footnote{\url{https://en.wikipedia.org/wiki/Wikipedia:List\_of\_spelling\_variants}} of the keywords to allow for difference in word usage across English dialects.
  \item We combined the search keywords, their corresponding synonyms, truncated forms, and alternatives spelling using the boolean \texttt{OR} operator.
  \item We then used the boolean \texttt{AND} operator to combine the search strings (combination of search keywords, their synonyms, truncated forms, and alternative spellings) corresponding to population, intervention, and outcome.
\end{enumerate}

The search strings obtained by applying the aforementioned approach are as follows.

\begin{description}
  \item[Population] \hfill
  
  \begin{itemize}
    \item ``software'' \llor ``application'' \llor ``system'' \llor ``program'' \llor ``product'' \llor ``code''
  \end{itemize}
  
  \item[Intervention] \hfill
  
  \begin{itemize}
    \item ``metric'' \llor (``measure'' \llor ``measurement'')
  \end{itemize}

  \item[Outcome] \hfill
  
  \begin{itemize}
  	\item ((``discover'' \llor ``detect'' \llor ``predict'' \llor ``uncover'' \llor ``locate'') \lland (``vulnerability'' \llor ``vulnerabilities'' \llor ``security vulnerability'' \llor ``security vulnerabilities''))
  	\item ((``vulnerability'' \llor ``security vulnerability'') \lland (``discovery'' \llor ``detection'' \llor ``prediction''))
  \end{itemize}
  
  \item[Context] \hfill
  
  \begin{itemize}
    \item ``quantitative'' \llor ``empirical'' \llor ``evidence-based'' \llor ``experiential''
  \end{itemize}
\end{description}

The individual search strings corresponding to the population, intervention, outcome, and context are logically combined to using the form (outcome \lland intervention \lland population \lland context). The effective search string is shown in Figure \ref{fig:searchstring}

\begin{figure}[ht]
  \centering
  \begin{framed}
  	(((``discover'' \llor ``detect'' \llor ``predict'' \llor ``uncover'' \llor ``locate'') \lland (``vulnerability'' \llor ``vulnerabilities'' \llor ``security vulnerability'' \llor ``security vulnerabilities'')) \llor ((``vulnerability'' \llor ``security vulnerability'') \lland (``discovery'' \llor ``detection'' \llor ``prediction''))) \lland (``metric'' \llor (``measure'' \llor ``measurement'')) \lland (``software'' \llor ``application'' \llor ``system'' \llor ``program'' \llor ``product'' \llor ``code'') \lland (``quantitative'' \llor ``empirical'' \llor ``evidence-based'' \llor ``experiential'')
  \end{framed}
  \caption{Effective search string obtained by logically combining the individual search strings determined for population, intervention, outcome, and context of interest in the systematic review}
  \label{fig:searchstring}
\end{figure}

\header{Inclusion and Exclusion Criteria Select Primary Studies} % -----------------------------

By applying the search strategy, we will identify candidate studies that are hypothesized to be relevant to our systematic review. However, not all candidate studies may contain the data needed to address all the research questions posed in our systematic review. In this step of the review process, the candidate studies are subject to additional criteria (referred to as inclusion and exclusion criteria) to identify those studies that are relevant to our review. The inclusion and exclusion criteria used in our systematic review are described in the subsections that follow.

Applying the inclusion and exclusion criteria to evaluate the relevancy of candidate studies, for the most past, an objective exercise. As a result, the benefit of having at least two authors independently apply the inclusion and exclusion criteria to all the candidate studies may be trivial. However, to quantify the objectivity of applying the inclusion and exclusion criteria, a random subset of the candidate studies will be independently evaluated for relevancy by at least two authors. The level of agreement between the two authors will be quantified using the Cohen's $\kappa$. If the level of agreement is not almost perfect (See scale in a paper by Landis and Koch \cite{Landis1997Measurement}), all candidate studies will be independently evaluated for relevancy by at least two authors to mitigate subjectivity. In the event that two or more authors apply the inclusion and exclusion criteria to all candidate studies, the level of agreement between the authors will be assessed using Cohen's $\kappa$, if two authors were involved, or Fleiss' $\kappa$, if more than two authors were involved. Any disagreements in applying the inclusion and exclusion criteria will be resolved through discussion among the authors involved. If the discussion yields no consensus, an additional author may be involved to mediate the disagreement.

\subheader{Exclusion Criteria}

The exclusion criteria defines rules to exclude a candidate study from consideration. The exclusion criteria in our review is as follows.

\begin{itemize}
  \item Studies for which the published full text is inaccessible through the University's subscription.
  \item Studies not written in English.
  \item Studies that have not been subject to peer review.
  \item Studies published as extended abstracts or supplement to poster and/or presentation.
  \item Studies that are secondary or tertiary.
  \item Studies published before the year 2000.\footnote{The year threshold is based on a similar study by \etal{Morrison} \cite{Morrison2018Mapping}. The threshold, while arbitrary, is reasonable since the term \textit{vulnerability} was formally defined by Krsul in his PhD thesis \cite{Krsul1998Software} in 1998.}
  \item Studies not published in a venue related to the discipline of Computer Science.
\end{itemize}

\subheader{Inclusion Criteria}

\begin{itemize}
  \item Study proposes and/or evaluates one or more metrics as a means to discover vulnerabilities in software.
  \item Study presents empirical evidence when reasoning about the utility of the metrics to discover vulnerabilities in software.
  
  The empirical evidence presented is in the context of publicly-disclosed historical vulnerabilities either disclosed via the National Vulnerability Database (NVD) or as advisories or bulletins in vendor-specific security disclosure portals such as Microsoft Security Bulletin\footnote{\url{https://technet.microsoft.com/en-us/security/bulletins.aspx}} and Mozilla Foundation security Advisories\footnote{\url{https://www.mozilla.org/en-US/security/advisories/}}. The rationale behind this qualification is that we do not want to include studies that \emph{assume} static analysis warnings to be indicative of real vulnerabilities and propose and/or evaluate metrics to discover static analysis warnings.
  \item Study describes the approach to collect the metric being proposed and/or evaluated with enough detail to enable replication.
  \item Study provides an interpretation of the metric as being a factor in vulnerability discovery.
  \item Study informs the practice of Software Engineering.
\end{itemize}

\header{Primary Study Quality Assessment} % -----------------------------

Assessment of the quality of a primary study is necessary to assess the extent to which a primary study minimized bias and maximized internal and external validity. The quality assessment is usually guided by an quality instrument that takes the form of a questionnaire. The questions in the quality assessment instrument may apply to different stages (design, conduct, analysis, conclusions, and reporting) of the research study being assessed.

The guidelines \cite{Kitchenham2007Guidelines} prescribe the use of quality assessment as a means to \begin{enumerate*}[label=(\alph*)] \item assist in primary study selection wherein the quality assessment instrument may be used to inform the development of the inclusion and exclusion criteria or \item assist data analysis and synthesis wherein the quality assessment instrument can be used to inform the analysis of the data extracted from the primary studies\end{enumerate*}.

In our review, we will not use the quality assessment to exclude certain primary studies because our intention is to get as eclectic a view about vulnerability discovery as possible but rather use one to quantify the quality of a primary study. Furthermore, we will use the quality of primary studies so quantified merely as a reporting metric.

The quality assessment instrument used in our systematic review is shown in Table \ref{atab:qualityassessmentinstrument} in Appendix \ref{asec:qualityassessmentinstrument}. The questions included in the quality assessment instrument are based on the quality checklist presented in Table 5 of the guidelines \cite{Kitchenham2007Guidelines}. Each question in the quality assessment instrument may be answered with a \emph{yes}, \emph{no}, or \emph{partially}. Each possible answer choice is assigned a numerical value with \emph{yes} being 1.0, \emph{no} being 0.0, and \emph{partially} being 0.5. The numerical value of the answer to each of the quality assessment questions is added up giving each primary study an overall quality assessment score. We will report this score when presenting synthesizing data from the corresponding primary study.

\header{Strategy for Data Extraction} % -----------------------------

The data needed to address the research questions addressed in our systematic review will be extracted from the primary studies using a predetermined data extraction form. The design of the data extraction form is driven by the research questions and the type of data needed to address the research questions. The data extraction form that will be used in our systematic review is presented in Table \ref{atab:dataextrationform} in Appendix \ref{asec:dataextractionform}.

The process of extracting data from primary studies, for the most part, is an objective exercise. As a result, the benefit of having at least two authors independently extract the same data from all primary studies may be trivial. However, to quantify the objectivity of the data extraction process, at least two authors will independently extract data from a random subset of the primary studies. The level of agreement between the two authors will be quantified using the Cohen's $\kappa$. If the level of agreement is not almost perfect (See scale in a paper by Landis and Koch \cite{Landis1997Measurement}), at least two authors will independently extract data from all primary studies to mitigate subjectivity. In the event that two or more authors extract the data from all primary studies, the level of agreement between the authors will be assessed using Cohen's $\kappa$, if two authors were involved, or Fleiss' $\kappa$, if more than two authors were involved. Any disagreements in data extraction will be resolved through discussion among the authors involved. If the discussion yields no consensus, an additional author may be involved to mediate the disagreement.

\header{Approach to Synthesize Extracted Data} % -----------------------------

There are two ways of synthesizing data for addressing research questions in a systematic review: descriptive (narrative) synthesis and quantitative synthesis. Although primary studies in our systematic review are likely to be empirical in nature, using formal meta-analysis to synthesize data in a quantitative way may be infeasible since the protocol for reporting quantitative results tend to vary greatly between primary studies \cite{Brereton2007Lessons}. As a result, we will use the descriptive (narrative) synthesis approach to synthesize the data needed for addressing the research questions in our systematic review.

\subsubsection{Protocol Evaluation}

The review protocol is perhaps the most important factor in ensuring that the systematic review is carried out in as rigorous, unbiased, and repeatable way as possible. The review protocol detailed in this report has been collaboratively developed by two authors to uncover any methodological flaws as early as possible. Additionally, the protocol will be reviewed by a third author to mitigate any biases that the protocol authors may have induced. We will also use AMSTAR (A MeaSurement Tool to Assess systematic Reviews) \cite{Shea2017AMSTAR} to assess the quality of the review protocol. The final step in evaluating the protocol is to pilot the entire systematic review in accordance with the procedures outlined in the protocol for a subset of candidate studies.

\appendix

\section{Quasi-Gold Standard Set}\label{asec:qgsset}

The primary studies listed below comprise the Quasi-Gold Standard set that was used to validate the search string in accordance with the Quasi-Gold Standard approach prescribed by \etal{Zhang} \cite{Zhang2011Identifying}. The primary studies that comprise the Quasi-Gold Standard set were curated based on personal experience of the authors in conducting empirical research in vulnerability discovery metrics. The evidence of this fact is in five of the primary studies in the Quasi-Gold Standard set being authored by at least one of the authors of this systematic review.

\begin{enumerate}[label=QGS\arabic*]
  \item \label{item:Neuhaus2007Predicting} Predicting Vulnerable Software Components \cite{Neuhaus2007Predicting}
  \item \label{item:Shin2008Empirical} An Empirical Model to Predict Security Vulnerabilities Using Code Complexity Metrics \cite{Shin2008Empirical}
  \item \label{item:Neuhaus2009Beauty} The Beauty and the Beast: Vulnerabilities in Red Hat's Packages \cite{Neuhaus2009Beauty}
  \item \label{item:Meneely2010Strengthening} Strengthening the Empirical Analysis of the Relationship Between Linus' Law and Software Security \cite{Meneely2010Strengthening}
  \item \label{item:Zimmermann2010Searching} Searching for a Needle in a Haystack: Predicting Security Vulnerabilities for Windows Vista \cite{Zimmermann2010Searching}
  \item \label{item:Shin2011Evaluating} Evaluating Complexity, Code Churn, and Developer Activity Metrics as Indicators of Software Vulnerabilities \cite{Shin2011Evaluating}
  \item \label{item:Shin2011Initial} An Initial Study on the Use of Execution Complexity Metrics As Indicators of Software Vulnerabilities \cite{Shin2011Initial}
  \item \label{item:Chowdhury2011Using} Using complexity, coupling, and cohesion metrics as early indicators of vulnerabilities \cite{Chowdhury2011Using}
  \item \label{item:Shin2013Traditional} Can traditional fault prediction models be used for vulnerability prediction? \cite{Shin2013Traditional}
  \item \label{item:Meneely2013Patch} When a Patch Goes Bad: Exploring the Properties of Vulnerability-Contributing Commits \cite{Meneely2013Patch}
  \item \label{item:Scandariato2014Predicting} Predicting Vulnerable Software Components via Text Mining \cite{Scandariato2014Predicting}
  \item \label{item:Walden2014Predicting} Predicting Vulnerable Components: Software Metrics vs Text Mining \cite{Walden2014Predicting}
  \item \label{item:Perl2015Finding} VCCFinder: Finding Potential Vulnerabilities in Open-Source Projects to Assist Code Audits \cite{Perl2015Finding}
  \item \label{item:Younis2016Fear} To Fear or Not to Fear That is the Question: Code Characteristics of a Vulnerable Function with an Existing Exploit \cite{Younis2016Fear}
  \item \label{item:Munaiah2016Beyond} Beyond the Attack Surface: Assessing Security Risk with Random Walks on Call Graphs \cite{Munaiah2016Beyond}
  \item \label{item:Munaiah2017Foreshadow} Do bugs foreshadow vulnerabilities? An in-depth study of the Chromium project \cite{Munaiah2017Foreshadow}
  \item \label{item:Munaiah2017Natural} Natural Language Insights from Code Reviews that Missed a Vulnerability \cite{Munaiah2017Natural}
\end{enumerate}

\section{Quality Assessment Instrument}\label{asec:qualityassessmentinstrument}

The questionnaire that will be used to assess the quality of primary studies included in the systematic review is shown in \ref{atab:qualityassessmentinstrument}.

\begin{table}[ht]
\renewcommand{\arraystretch}{1.3}
\centering
\caption{Questionnaire used to assess the quality of primary studies included in the systematic review}
\label{atab:qualityassessmentinstrument}
\begin{tabular}{L{0.10\textwidth}L{0.80\textwidth}}
  \hline
  \ccb{Stage}        & \ccb{Question} \\
  \hline
  \mr{4}{Design}     & Does the study have a clearly-stated aim or goal? \\ \cdline{2-2}
                     & Do the research questions address the aim or goal of the study? \\ \cdline{2-2}
                     & Is the choice of subject of study in the emprical analysis justified? \\ \cdline{2-2}
                     & Are the metrics used to address the research questions adequately described? \\ \hline
  \mr{3}{Conduct}    & Has the problem of class imbalance addressed? \\ \cdline{2-2}
                     & If predictability of the metrics was considered, how was the model validated? \\ \cdline{2-2}
                     & Are the data types of the metric reported? \\ \hline
  \mr{7}{Analysis}   & Were confounding factors of a metric considered? \\ \cdline{2-2}
                     & Were descriptive statistics of the metrics reported? \\ \cdline{2-2}
                     & Are the statistical methods adequately described? \\ \cdline{2-2}
                     & Are the statistical software/packages used referenced? \\ \cdline{2-2}
                     & Are the statistical methods used adequately justified? \\ \cdline{2-2}
                     & Is the actual p-value reported? \\ \cdline{2-2}
                     & Was effect size reported? \\ \hline
  \mr{4}{Conclusion} & Were all research questions answered? \\ \cdline{2-2}
                     & Were the implications of the study described? \\ \cdline{2-2}
                     & Were limitations and/or threats to validity reported? \\ \cdline{2-2}
                     & Were the findings compared with findings from prior literature? \\
  \hline
\end{tabular}
\end{table}

\section{Data Extraction Form}\label{asec:dataextractionform}

The form that will facilitate the extraction of data from primary studies to address the research questions is shown in Table \ref{atab:dataextrationform}.

\begin{center}
\renewcommand{\arraystretch}{1.3}
\begin{longtable}{L{0.20\linewidth}L{0.20\linewidth}L{0.55\linewidth}}
  \caption{Data extraction form}\label{atab:dataextrationform}\\
  % First Header
  \hline \ccb{Item} & \ccb{Data Type}& \ccb{Description} \\ \hline
  \endfirsthead

  % Repeated Header
  \multicolumn{3}{C{\linewidth}}%
  {{\tablename\ \thetable{}. Data extraction form (Continued)}} \\
  \hline \ccb{Item} & \ccb{Data Type}& \ccb{Description} \\ \hline
  \endhead

  % Repeated Footer
  \multicolumn{3}{r}{{Continued on next page}} \\
  \endfoot

  % Last Footer
  \hline
  \endlastfoot

% Table Rows
\mcb{3}{c}{\cellcolor{gray!15}Miscellaneous Metadata}\\ \hline
Extracted By              & Plain Text  & The name of the researcher who extracted the data. \\ \dline
Verified By               & Plain Text  & The name of the researcher who verified the extracted data. \\ \dline
Identifier                & Plain Text  & The unique identifier of the primary study from which the data was extracted. \\ \dline
\# Subjects               & Numeric     & The number of subjects of study that a metric have been collected from. \\ \dline
Subjects' Nature          & Categorical & The type of subject of study. \\ \dline
Subjects' Domain          & Plain Text  & The application domain of the subject of study. \\ \dline
Response Variable         & Plain Text  & The description of the response variable in the study. In some vulnerability discovery studies, we have seen the Number of Vulnerabilities being the response variable rather than Vulnerableness. The response variable is an important piece of information that must be captured as part of the data extraction. \\ \dline
\# Vulnerabilities        & Numeric     & The number of vulnerabilities that were considered in the empirical analysis. \\ \dline
Miscellaneous             & Varying     & Miscellaneous metadata about the primary study such as year of publication, name of the journal/conference/workshop in which the study was included, and number of pages. \\ \hline

\mcb{3}{c}{\cellcolor{gray!15}RQ1 Enumeration}\\ \hline
Name                      & Plain Text  & The name of a vulnerability discovery metric. \\ \dline
Definition                & Plain Text  & The definition of a vulnerability discovery metric. \\ \dline
Unit                      & Plain Text  & The unit of the vulnerability discovery metric. \\ \dline
Applicability             & Categorical & The applicability of the metric. For instance, if the Number of Late-night Commits is posited to be related to vulnerability discovery, the applicability of the metric would be categorized as Process because the aspect of Software Engineering that the metric applies to is Process (Version Control) \\ \dline
Granularity               & Plain Text  & If applicable, the granualarity at which the metric is defined. For instance, if SLOC is posited to be related to vulnerability discovery, was SLOC collected at the function-level, method-level, class-level, file-level, or component-/module-level. \\ \hline

\mcb{3}{c}{\cellcolor{gray!15}RQ2 Validation}\\ \hline
Validation Criteria       & Categorical & The name of the validation criteria that a metric has been subject to. The list of 47 validation criteria from Validating Software Metrics: A Spectrum of Philosophies by Meneely et al. will be used. \\ \dline
Validation Approach       & Plain Text  & A metric may subject to one of more validation criteria and the test for each validation criteria may be different across studies. For instance, the "Association" of a metric may be validated using a simple correlation analysis using Spearman's, an association analysis using t-test or MWW, or by analyzing the coefficient of the metric in a regression model. \\ \dline
Evaluation Metrics        & Plain Text  & When assessing the validity of a metric using a particular test, there may be a variety of metrics that could be used. For instance, when assessing the predictability of metrics using a classifier, one can use a variety of metrics such as precision, recall, f-measure, TPR, and FPR. We should capture these metrics to report. \\ \dline
Evaluation Metrics' Value & Numeric     & For each evaluation metric used to validate a metric, the raw value of the metric as reported. For instance, if improvement in precision, recall, or f-measure was a validition metric used, then the raw value of percentage improvement would be captured. \\ \dline
Was Aggregated            & Boolean     & In assessing the validity of a metric, were the metric values aggregated to a higher granularity at which at which they were collected? \\ \dline
Aggregation Rationale     & Plain Text  & If a metric was aggregated for validation, describe the rationale for aggregation as presented by the researchers. For instance, if Churn is posited to be related to vulnerability discovery, researchers are likely to collect churn at the commit-level and then aggregate it to the file-level for analysis. The rationale for analyzing Churn at the file-level is that it can be used to reason about the amount of change a file has been through. One possible interpretation of high churn in a file is that it increases the likelihood of changes being not thoroughly reviewed during code review. \\ \hline

\mcb{3}{c}{\cellcolor{gray!15}RQ3 Applicability}\\ \hline
Metric Interpretation     & Plain Text  & The description of the interpretation of the metric as reasoned by the authors. \\ \dline
Metric Control            & Plain Text  & If available, the description of the way in which developers may use the metric to lower the risk of vulnerability. For instance, if Number of Late-night Commits is posited to be related to vulnerability discovery, one of the ways in which a researcher could propose using the metric is to enforce a thorough code review of late-night commits. \\ \hline

\mcb{3}{c}{\cellcolor{gray!15}RQ4 Prerequisites}\\ \hline
Metric Prerequisites      & Plain Text  & A list of prerequisites that must be satisfied to be able to collect a specific metric from a subject of study. For instance, if the location of the developer is a metric posited to be related to vulnerability discovery, the ability to collect developers' location information is a prerequisite. Similarly, for the Known Offender metric, having historical vulnerability information is a pre-requisite. \\ \dline
Approach                  & Plain Text  & A description of the approach to collect a metric as described by the authors. \\ \dline
Metric Collection Time    & Numeric     & The time taken to collect a metric from a software system. Assuming the pre-requisites of a metric are satistifed, if the time taken to collect a metric is prohibitive, developers are less likely to collect the metric and use the intelligence that the metric may afford. \\ \dline
Metric Limitations        & Plain Text  & A description of any limitations that may apply to a metric. For instance, if Afferent or Efferent Coupling was posited to be related to vulnerability discovery, the limitation of the metric would be that it applied only to object-oriented languages. \\

\end{longtable}
\end{center}

\bibliographystyle{plain}

\end{document}